\documentstyle[prb,aps,epsfig,multicol]{revtex}

\begin{document}
\title{Effects of C, Cu and Be Substitutions in Superconducting MgB$_2$}
\author{M.J. Mehl, D.A. Papaconstantopoulos and D.J. Singh}
\address{Center for Computational Materials Science,\\
Naval Research Laboratory, Washington, DC 20375, U.S.A.}
\date{\today}
\maketitle

\begin{abstract}
Density functional calculations are used to investigate the effects
of partial substitutional alloying of the B site in MgB$_2$ with
C and Be alone and combined with alloying of the Mg site with Cu.
The effect of such substitutions on the electronic structure, electron
phonon coupling and superconductivity are discussed. We find
that Be substitution for B is unfavorable for superconductivity
as it leads to a softer lattice and weaker electron-phonon couplings.
Replacement of Mg by Cu leads to an increase in the stiffness and
doping level at the same time, while the carrier concentration can
be controlled by partial replacement of B by C.
We estimate that with full replacement of Mg by Cu
and fractional substitution of B by
C, $T_c$ values of 50K may be attainable.
\end{abstract}

\begin{multicols}{2}


The discovery of superconductivity with critical temperature 
$T_c$ = 39 K in hexagonal MgB$_2$ has led
to considerable interest in this material, both for practical
applications and from a more fundamental
point of view. \cite{akimitsu} In particular, although it is
not the highest $T_c$ conventional (in the sense of $s$-wave
with likely substantial electron-phonon coupling) superconductor, \cite{schon}
its substantial $T_c$ combined with its chemistry and other
properties may make it particularly useful.
\cite{can1,larb,bug}
A variety of experimental measurements, including B isotope effect,
\cite{budko}
neutron scattering phonon densities of states,
\cite{sato,osborn}
tunnelling, \cite{sharoni}
transport, \cite{can1,finnemore,budko2}
and specific heat, \cite{kremer,walti}
have been done. Taken together, they yield a consistent
clear picture of MgB$_2$ as a conventional electron-phonon $s$-wave
superconductor. This is also consistent with most of the reported
theoretical work.

The underlying electronic structure and bonding were discussed by
Ripplinger \cite{ripplinger}
before the discovery of superconductivity. This was based on
density
functional calculations using the linearized augmented planewave
(LAPW) method and provided results consistent with an early
calculation by Armstrong and Perkins. \cite{armstrong}
Subsequently, several authors have extended this
work presenting the electronic structure in detail and discussing
the origin of superconductivity.
\cite{an,kong,kortus,medvedeva}
The electronic structure is dominated by the
chemical bonding of the hexagonal B sheets in MgB$_2$.
Although the nominal electron count is the same as in graphite (Mg has
its nominal 2+ charge), unlike graphite,
the top of the B derived bonding band structure
contains holes related to fractional occupation of 
bands derived from B $p$ states directed perpendicular to the layers.
As pointed out by An and Pickett, \cite{an}
the negative charge of the B layers and the corresponding positive
charge of the Mg layers leads to a lowering of the $p_z$
bands relative to their position in graphite.
Detailed calculations of the phonon spectra and electron phonon
coupling have been reported.
These show strong coupling between
the hole "doped" bonding bands and high frequency optical phonons
associated with motions of the B atoms affecting the covalent bonds.
Kong {\it et al.} reported detailed calculations of electron-phonon
couplings over the whole Brillouin zone obtaining couplings consistent
with the measured $T_c$ as well as the boron isotope effect,
the specific heat enhancement, the reported gap values and
transport data. \cite{kong}

The picture that emerges is one where the high $T_c$
is due to strong electron phonon coupling associated with the hole doped
metallic $\sigma$ bonding bands in the B sheets. This is due to the strong
bonding of these sheets, which along with the light B mass is responsible
for the high average frequency of the strongly coupled phonons,
setting the temperature scale. However, there are also significant
interactions with lower frequency modes and this may be needed to explain
the upper critical field. \cite{kong,shugla}
Though some interesting questions remain, {\it e.g.}, regarding
possible anharmonic effects for the lower frequency phonon modes,
\cite{shugla,liu}
the basic physics appears to be settled as regards the origin of
superconductivity in MgB$_2$.
The crucial aspects seem to be: (1) band structure, particularly
the presence of hole doped $\sigma$ bands at the Fermi level, $E_F$,
(2) strong electron phonon coupling associated with the strong
covalent bonding nature of these bands and (3) high phonon frequencies
associated again with the strong covalent bonds and the light B mass.

The present work is to explore some possible substitutions
and provide suggestions for experimental work aimed at finding
related high temperature superconducting phases. Along these lines,
Medvedeva {\it et al.} have investigated a number of possible
substitutions of Mg by monovalent, divalent and trivalent ions.
\cite{medvedeva}
They focussed on the band structure, and particularly the density
of states (DOS) and presence or absense of the $\sigma$ band at $E_F$
They concluded that trivalent substitutions like Y, Al etc. are not
favorable as they fill the hole doped (in MgB$_2$) $\sigma$ bands,
while certain monovalent substitutions for Mg may be favorable.
This is consistent with the experimental observation that Al substitution
destroys superconductivity. \cite{slusky}
They also mentioned the possibility of vacancies or substitutions
on the B sheets, but concluded that these are all unfavorable.

Here we briefly reexamine the effect of Be substitution on the B planes,
including band structure, lattice stiffness and electron-phonon
calculations. Although Be substitution lowers the electron count,
we find that it is so detrimental to the bonding that superconductivity
is suppressed both due to a strong decrease in the lattice
stiffness and a drop in the electron-phonon coupling. We then report
calculations investigating the effect of a combined
partial substitution of
C for B and Cu for Mg. The rationale for this combination is that
C-B bonds are expected to be very strong in this structure, and so
this substitution may lead to a stiffening of the sheets
relative to MgB$_2$, while the replacement of Mg by Cu may be expected
to first of all compensate for the extra charge provided by the substitution
in the plane, and secondly, perhaps maintain the hole doping of the $\sigma$
bonding band, present in MgB$_2$ but absent in graphite. The basis of
this latter conjecture is that the crystal field in materials
can arise both from strict ionic (Coulomb) effects, as seem to dominate
in MgB$_2$, and hybridization, which is important {\it e.g.} in
transition metal oxides. The hope then was that such effects with monovalent
Cu might be sufficient to keep the $\sigma$ band at the Fermi energy.
As discussed below we find that both these effects do occur and that
for modest C incorportation and Cu replacement of Mg superconductivity
should be enhanced. Based on rigid muffin tin approximation (RMTA)
calculations, we estimate that $T_c$ may be near 50K.

\begin{figure}[tbp]
\centerline{\epsfig{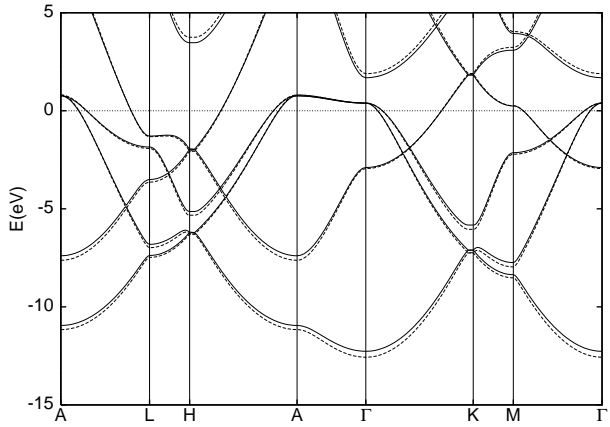}}
\vspace{0.125in}
\setlength{\columnwidth}{3.2in} \nopagebreak
\caption{
LDA band structure of MgB$_2$ with the experimental structure (solid)
and calculated LDA lattice parameters (dashed). The zero is at $E_F$.
}
\label{ref-bands}
\end{figure}

Our results were obtained in the local density approximation (LDA)
\cite{hl} using the general potential LAPW method, including local orbitals,
\cite{singh-book,singh-lo} and well converged basis sets and zone
samplings. Band structures, lattice stiffnesses and electron phonon
couplings were all determined at the calculated lattice parameters as
determined by energy minimization.
Commonly in LDA calculations, lattice constants are slightly underestimated.
For MgB$_2$ we get $a$=5.736 $a_0$ and $c$=6.522.$a_0$,
as compared to the experimental values $a$ =5.826 $a_0$ and $c$=6.653 $a_0$.
This is a 5\% volume compression. As shown in Fig. \ref{ref-bands}
this has a very small effect on the band structure near the
Fermi energy, $E_F$ with a small effect on $N(E_F)$.
To proceed to the alloys and discuss
superconductivity, we make three more significant approximations. First of
all, we employ the virtual crystal approximation (VCA) to account for partial
substitutions on the B sheets. A partial justification for this is
provided by the strong covalency and corresponding large bandwidths,
which may limit the amount of scattering due to potential disorder
in the alloy (this is the same effect that allows alloys,
like Al$_x$Ga$_{1-x}$As, to have high enough mobilities to be useful in
semiconductor technology). We tested the VCA by
comparing with ordered cells at the compositions MgBeB and CuBC and
found some quantitative differences (discussed below), but the key
features of the band shapes, velocities and the position of the
$\sigma$ bonding band relative to $E_F$ were little changed.
The second approximation we made was to characterize the lattice
stiffness that combined with the mass gives average phonon frequencies
by the calculated tensile stiffness of the B sheets, {\it i.e.}
$\partial^2 E/\partial a^2$. This number and the average
mass of the atoms in the B sheets was used to scale the
average frequencies as calculated by {Kong \it et al.} \cite{kong}
Considering that the dominant phonons
determining $T_c$ are the B modes, we think that this is
a reasonable approximation. Finally, we use the RMTA
to characterize the electron phonon coupling.
\cite{rmta1,rmta2}

We used well converged DOS calculations based on first
principles eigenvalues at over 2000 {\bf k} points in the
irreducible zone to get $N(E_F)$ and the corresponding
angular momentum components.  We used the self-consistent LAPW
potentials at different concentrations to calculate the corresponding phase
shifts and free scatterer DOS at $E_F$.
Sphere radii of 1.5 $a_0$
were used for B, Be and C for the RMTA calculations.
The above quantities were used in the
Gaspari-Gyorffy formula to compute the Hopfield  parameters, $\eta$,
for each site.
Negligible coupling is found on the Mg site, as expected, but not
Cu, {\it e.g.} 0.78 eV/\AA$^2$ on Cu for CuB$_2$.
However, in Table \ref{tab-data}
we give the values of $\eta$ for B only,
since that is the dominant contribution controlling superconductivity
and we do not include any Cu contribution in the calculation of $\lambda$
or $T_c$
(Cu will have little
involvement in the high frequency phonons associated with the B sheets).
For the electron-phonon coupling we used the usual expression
$\lambda=\eta/{<M\omega^2>}$.
For
the denominator we used the average frequency calculated by Kong {\it et al.}
\cite{kong}
for MgB$_2$ and scaled it using the
tensile stiffness of the B sheets for various concentrations,
{\it i.e.}  $\partial^2E/\partial a^2$.
The RMTA is not generally as well justified in $sp$ metals
as in transition metals and can considerably underestimate
the deformation potentials when strong $sp$ covalent bonding is present as
it is here.
Further, the RMTA
neglects some differences between different bands, which may be
significant here. In any case, Kortus {\it et al.} did use it
for MgB$_2$ to characterize electron phonon couplings.
\cite{kortus}
Comparing with the direct calculations of Kong {\it et al.} for MgB$_2$
we find, not unexpectedly, that the values of $\eta$ we obtain with
our non-overlapping B spheres are too small, roughly by a factor of three.
Here we use
the RMTA to elucidate trends, while acknowledging its limitations.
So we adopt the heuristic of scaling the calculated values of $\eta$
by three in calculating $\lambda$ and $T_c$. Using the average
phonon frequencies quoted by Kong {\it et al.} this heuristic
closely reproduces their values of $\lambda$ and $T_c$ for MgB$_2$.
We used the McMillan equation to roughly estimate $T_c$, setting
the Coulomb pseudopotential $\mu^*$=0.1.
We emphasize here that we are not aiming at an accurate determination of
the value of $T_c$ but exploring the trends upon substituting Mg by Cu
and B by Be or C. Because of our use of the RMTA our values
should be taken as indicating trends not as quantitative 
predictions of $T_c$.

\begin{figure}[tbp]
\centerline{\epsfig{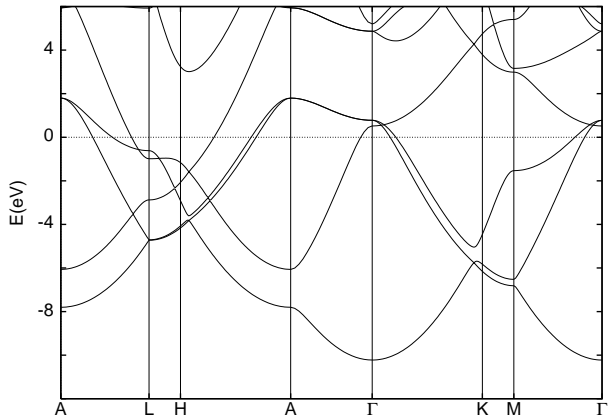}}
\vspace{0.125in}
\setlength{\columnwidth}{3.2in} \nopagebreak
\caption{
LDA virtual crystal band structure of MgBe$_x$B$_{2-x}$ for $x=1$.
The lattice parameters is
the calculated relaxed value.
The horizontal reference at 0 denotes $E_F$.
}
\label{be-bands}
\end{figure}
\begin{figure}[tbp]
\centerline{\epsfig{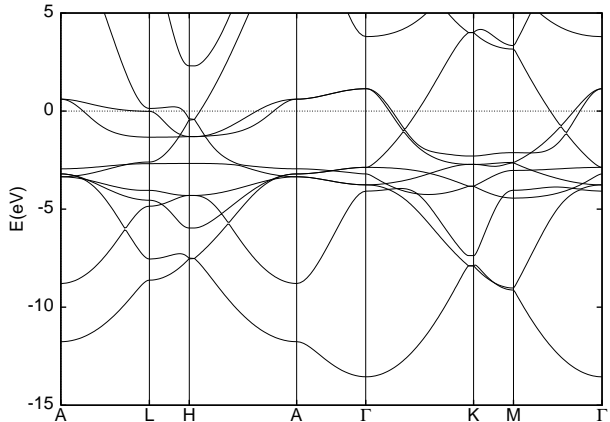}}
\vspace{0.10in}
\centerline{\epsfig{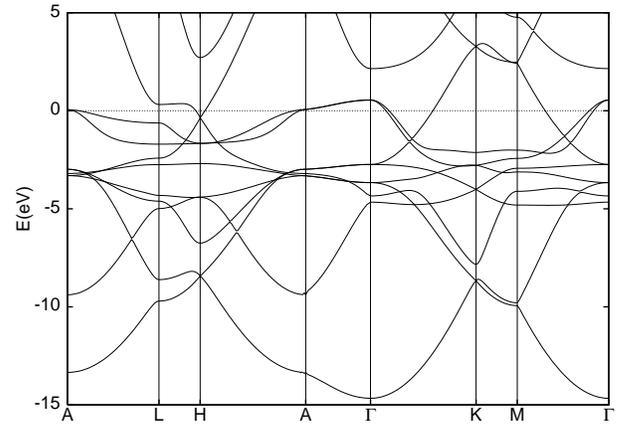}}
\vspace{0.10in}
\centerline{\epsfig{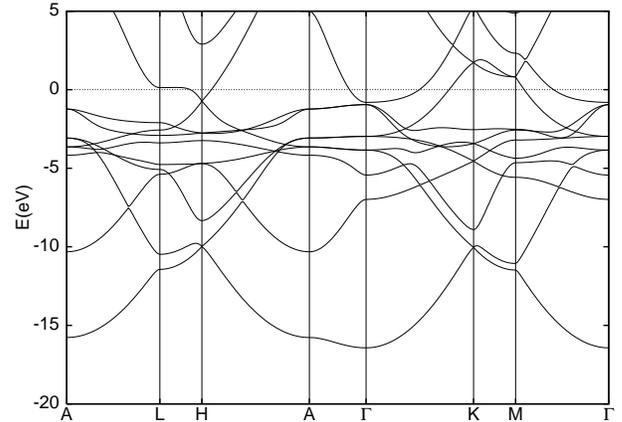}}
\vspace{0.125in}
\setlength{\columnwidth}{3.2in} \nopagebreak
\caption{
LDA virtual crystal band structures of CuB$_{2-x}$C$_x$ for $x=0$ (top),
$x=0.5$ (middle) and $x=1$ (bottom). The lattice parameters are
the calculated relaxed values. Note the vertical scale of the lower 
panel. The horizontal reference at 0 denotes $E_F$.
}
\label{cu-bands}
\end{figure}
\begin{figure}[tbp]
\centerline{\epsfig{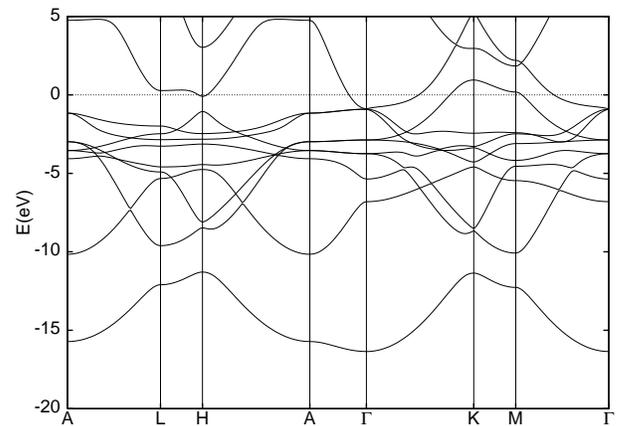}}
\vspace{0.125in}
\setlength{\columnwidth}{3.2in} \nopagebreak
\caption{
LDA ordered band structure for CuBC note the splittings
relative to the virtual crystal band structure in the
bottom panel of Fig. \ref{cu-bands}.
}
\label{cu-test}
\end{figure}

The band structure for the starting compound, MgB$_2$,
(Fig. \ref{ref-bands}) is practically identical to those obtained
previously, \cite{an,kong,kortus,medvedeva,notedos} showing
$\sigma$ bonding states at $E_F$. Results for the structural and
electronic properties relevant to superconductivity are given
in Table \ref{tab-data}, while band structures are shown in Fig. \ref{be-bands}
and Fig. \ref{cu-bands} for MgBe$_x$B$_{2-x}$ and CuB$_{2-x}$C$_x$,
respectively. In the table the values of $\eta$ are the bare values as
given by the RMTA, while the heuristically scaled values have
been used for the values of $\lambda$ and $T_c$.
An ordered band structure for CuBC is shown in Fig.
\ref{cu-test}. This shows some differences from the corresponding VCA
calculation, most notably, near $E_F$, a splitting at the H point
involving $p_z$ bands. However, the general structure of bands
near $E_F$ and the position of the $\sigma$ band is quite similar.

Substitution of Be into the B sheets lowers the electron count, though
as seen in the band structure, not in a rigid band way.
The hole concentration in the $\sigma$ bonding band does increase,
but the bonds are strongly weakened.
This is reflected in the band widths and lattice stiffness.
The result 
(Table \ref{tab-data}) is a rapid increase in the $a$ lattice
parameter, softening of the lattice and a decrease in the
electron phonon coupling. Thus Be substitution
is
detrimental to superconductivity and is not further discussed.

The Cu substituted material is more interesting.
In an ionic model, replacement of Mg by monovalent
Cu should lower the electron count in the B sheets by one per
formula unit. So perhaps a situation similar to that found
with 50\% Be substitution may be expected. This is not
the case. The Cu is monovalent as in the ionic model. The five
narrow Cu $d$ bands are in the valence region between -4 and -3 eV
relative to $E_F$.
Comparing the top panel of Figs. \ref{ref-bands} and
\ref{cu-bands} and the structural information in Table \ref{tab-data}
one sees that the bonding in the sheets is strengthened by the addition
of Cu, even though the hole count in the $\sigma$ band is
increased (note the relative positions of the band maxima on the
$\Gamma$ -- $A$ line). Hybridization with Cu leads to quantitative
changes in the band structure, affecting mostly the $p_z$ states,
but there are also weak effects on the $\sigma$ bands near $E_F$, {\it e.g.}
the reversed dispersion on the $\Gamma$ -- $A$ line presumably due
to $d$--$p\pi$ interactions. The net effects
of Cu substitution
-- stiffened lattice, increased $N(E_F)$ and higher hole concentration --
are favorable for superconductivity. Related to this, there is
a recent unconfirmed report of an enhancement of $T_c$ with partial
replacement of Mg by Cu.
\cite{sun}
Partial substitution of C for B in CuB$_2$ has two effects -- a stiffening
of the lattice reflecting the strong bonding of C and B (favorable
for $T_c$ and a reduction in the hole doping of the $\sigma$ band
and in $N(E_F)$ (unfavorable for $T_c$). For low C concentrations,
the first effect dominates, leading to an increase in the estimated $T_c$,
but beyond 25\% C substitution (here
the van Hove from the
$\sigma$ band at $A$ crosses $E_F$) the second dominates and $T_c$ falls.


In summary,
band structure and RMTA calculations indicate that the $T_c$
of MgB$_2$ can be increased, perhaps to 50K by substitution of Cu for 
Mg and low C substitutions, around 25\%, in the B sheets. We emphasize that
to our knowledge
these are not known equilibrium phases. However, the strong binding
of the sheets suggests synthesis may be possible, {\it e.g.} starting with
MgB$_2$ ({\it n.b.} Mg leaves the compound at high temperature).


This work was supported by ONR and the
the ASC supercomputer center. We are grateful for
discussions with I.I. Mazin, W.E. Pickett and K. Schwarz.

\begin{table}
\caption{Calculated properties of MgBe$_x$B$_{2-x}$ and 
CuB$_{2-x}$C$_x$ as obtained in the VCA.
$S$ is the relative stiffness of the B 
sheets characterized by $\partial^2E/\partial a^2$ ($S$(MgB$_2$)=1),
and the other symbols have their usual meanings. Lattice parameters
are in $a_0$, $N(E_F)$ in eV$^{-1}$ and $T_c$ in
K. The unscaled $\eta$ (in eV/\AA$^2$) is for the alloyed B site only.
Blank entries for $a$, $c$ and $S$ indicate that
these were obtained by interpolation rather than direct
computation for those concentrations.
}
\label{tab-data}
\vspace{0.125in}
\begin{tabular}{lddddddr}
                      &$a$ & $c$ &$S$ &$N(E_F)$& $\eta$ & $\lambda$ & $T_c$ \\
\tableline
MgB$_2$                & 5.74 & 6.52 & 1.00 & 0.68 & 3.60 & 0.78 & 38 \\
MgBe$_{0.5}$B$_{1.5}$  & 5.99 & 6.54 & 0.78 & 0.84 & 2.36 & 0.66 & 22 \\
MgBeB                  & 6.43 & 6.03 & 0.56 & 0.90 & 1.45 & 0.56 & 12 \\
MgBe$_{1.5}$B$_{0.5}$  & 6.90 & 5.53 & 0.48 & 0.98 & 1.04 & 0.47 & 6 \\
MgBe$_2$               & 7.32 & 5.34 & 0.47 & 0.95 & 0.93 & 0.43 & 4 \\

CuB$_2$                & 5.58 & 6.28 & 1.11 & 1.09 & 4.38 & 0.86 & 48 \\
CuB$_{1.75}$C$_{0.25}$ &      &      & 1.37 & 0.83 & 5.16 & 0.90 & 55 \\
CuB$_{1.5}$C$_{0.5}$   & 5.37 & 6.56 & 1.37 & 0.65 & 5.45 & 0.86 & 53 \\
CuB$_{1.4}$C$_{0.6}$   &      &      & 1.37 & 0.50 & 3.74 & 0.57 & 20 \\
CuB$_{1.25}$C$_{0.75}$ &      &      & 1.37 & 0.25 & 0.80 & 0.12 & 0  \\
CuBC                   & 5.11 & 7.28 & 1.63 & 0.40 & 1.02 & 0.14 & 0  \\
\end{tabular}
\end{table} 

\end{multicols}


\begin{references}
\bibitem{notedos}
Note that $N(E_F)$ in Table \ref{tab-data} is at the
LDA lattice parameters for consistency
and so is $\sim$ 4\% smaller
than obtained previously with experimental lattice paramters.
\end{references}
\end{document}